# Chapter N

# Modeling Responsible Elite

## Yana Tsodikova, Pavel Chebotarev, Anton Loginov, and Zoya Lezina

### V.A. Trapeznikov Institute of Control Sciences of RAS, Moscow

Within the ViSE (Voting in Stochastic Environment) model, we study social dynamics determined by collective decisions in a society with an elite. The model allows the analysis of the influence of participants' social attitudes, such as the effects of selfishness, collectivism, lobbying, altruism, etc., on the welfare of the society and its strata. Social dynamics is determined by the change in capital over time, as well as the formation and dissociation of groups. We show that the presence of a responsible elite, which partially cares for its own benefit, stabilizes society and removes the 'pit of damage' paradox. Society's gain from having a responsible elite is comparable to that from an altruistic group of the same size. If the responsible elite succumbs to the temptation to dramatically increase the weight of the group component in its combined voting strategy, then its income rises sharply, while the income of the society decreases. The rest of the participants benefit from joining the elite, while for the elite it is beneficial to maintain a moderate size. If, in response to insufficient responsibility of the elite, an altruistic group emerges that outnumbers the elite and becomes a new responsible elite, then society again stabilizes, and monopoly of the previous elite (a 'clique') ends. If the responsible elite competing with the clique becomes a second clique, then tough competition between the cliques is still preferable for society over having a unique clique.

## 1. Introduction

The ViSE (Voting in Stochastic Environment) model [1, 2], which has been investigated since 2003, is a model of social dynamics determined by collective decisions in a stochastic environment. The model allows the analysis of the influence of voting procedures used and participants' social attitudes (selfishness, collectivism, lobbying, altruism, etc.) on the welfare of the participants themselves, the whole society, and its strata.

In this model, each participant (agent) has its current *capital* (or *utility*), and the participants consecutively vote for the proposals generated by an external stochastic environment. Each proposal is a vector of changes in the capital of the agents; the components of this vector are supposed to be independent, identically distributed, random variables. The environment can be favorable or unfavorable, stable or volatile; these characteristics are modeled by the expectation (mean value) and variance of the random variables that generate the proposals. The type of distribution may vary. One can consider distributions with light or heavy tails, bell-shaped, or spiky, with limited or unlimited domain.

The vector of initial capitals of the agents determines initial conditions. Each agent evaluates a proposal based on the agent's voting strategy. For example, an *egoist* is interested in increasing their capital only, while a *collectivist* is interested in increasing the total capital of their group, and an *altruist* in increasing the capital of the whole society or, for example, its poorest stratum.

Collective decisions are made based on voting procedure. Accepted proposals are implemented, i.e., the capital of agents receives the prescribed increments. If a proposal is rejected, then the capital of the agents remains unchanged. The dependence of the capitals on time, as well as the formation and dissociation of factions, determine the social dynamics of society.

Thus, the ViSE model has a number of parameters, which can be interpreted in terms of applications. It allows one to build and analyze simple theoretical constructs reflecting the features of various homogeneous and heterogeneous societies. Some phenomena detected within the model can be observed, under certain circumstances, in reality. In addition, using the ViSE model, it is possible to find optimal voting procedures for societies with different structures in different environments. Moreover, this model enables one to find out the social attitudes of the agents that provide society with sustainable development. Finally, the model allows the design of optimal incentive schemes (for example, through the redistribution of income) for the participants whose social attitudes are valuable to society. The ViSE model has been analyzed in a number of papers, e.g., [1–6]. One of its significant differences from other comparable models [7–10] is that the focus of the research is transferred from the search for *equilibria* in legislative bargaining to the *maximization of utility* (capital), in other words, to *efficiency* in conditions that allow description in terms of games against nature. Efficiency analysis brings this approach closer to that implemented in [11], where there were no dynamics, and the number of proposals generated exogenously were put to vote simultaneously. However, the adoption of a proposal brought the agents certain capital/utility increments, and the dynamics of the capital increment vector allowed one to evaluate the effectiveness of the decision-making procedure. More on the comparison of the ViSE model with other related models can be found in [4–6].

## 2. Combined Voting Strategies

In this paper, we study *combined* voting strategies. The need for these is due to the following, previously noted, circumstances.

1. A society consisting of egoists is exposed to the danger of bankruptcy in a moderately unfavorable environment, due to the implementation of majority decisions (the so-called ‘pit of damage’ paradox [4]).
2. The presence of a relatively small faction of altruists protects the society from bankruptcy; however, altruists themselves turn out to be a low-income stratum.

A natural task is to provide agents having altruistic voting strategies with an acceptable income. One of the most interesting and realistic approaches to the solution of this problem is the use of *combined voting strategies*. An agent that combines altruistic and group strategies while evaluating a proposal computes a convex combination of the average increment of the capital of the whole society, and of the group they are a member of. If this convex combination is positive, they support the proposal, otherwise they vote against it.

A faction whose members adhere to such a combined voting strategy can be considered as a *responsible elite*. In this paper, we study societies that have such a faction and compare them to other societies.

## 3. Social Dynamics of a Society with a Responsible Elite

*3.1. A small number of altruists saves society from bankruptcy*

For definiteness, consider a society consisting of 201 agents. An odd number of participants leads to higher lability: a simple majority is achieved by an advantage of only one vote, with the result that proposals are accepted more often than with an even number of agents.

First, let all agents of the society be egoists: each of them votes for a proposal if, and only if, it provides him/her with an increase in the capital. The dynamics of society will be characterized by the mathematical expectation of the capital increment of a participant in one step (round) of voting (we will call it the *expected capital gain*, *ECG*).

This value is determined by the fraction of accepted proposals and the average increment of agent’s capital over accepted proposals. The ECG depends on the number of agents and the distribution that generates proposals, in particular, on the mean and variance of this distribution.

Fig. 1 shows the dependence of the ECG on the mathematical expectation $\mu$ of the normal distribution with standard deviation $\sigma = 34$ (the result was obtained analytically [2]; the subsequent curves are mainly obtained through simulation, since analytical expressions are hardly achievable). With moderate negative $\mu$, ECG takes noticeable negative values, i.e., society inevitably becomes poorer due to the implementation of majority decisions; the ECG curve has a ‘pit of damage’. This is because the majority approving a proposal receives on average less than the remaining minority loses.

This paradox can be overcome by varying the voting threshold [4] or changing the voting strategy. In this paper, we investigate the latter approach.

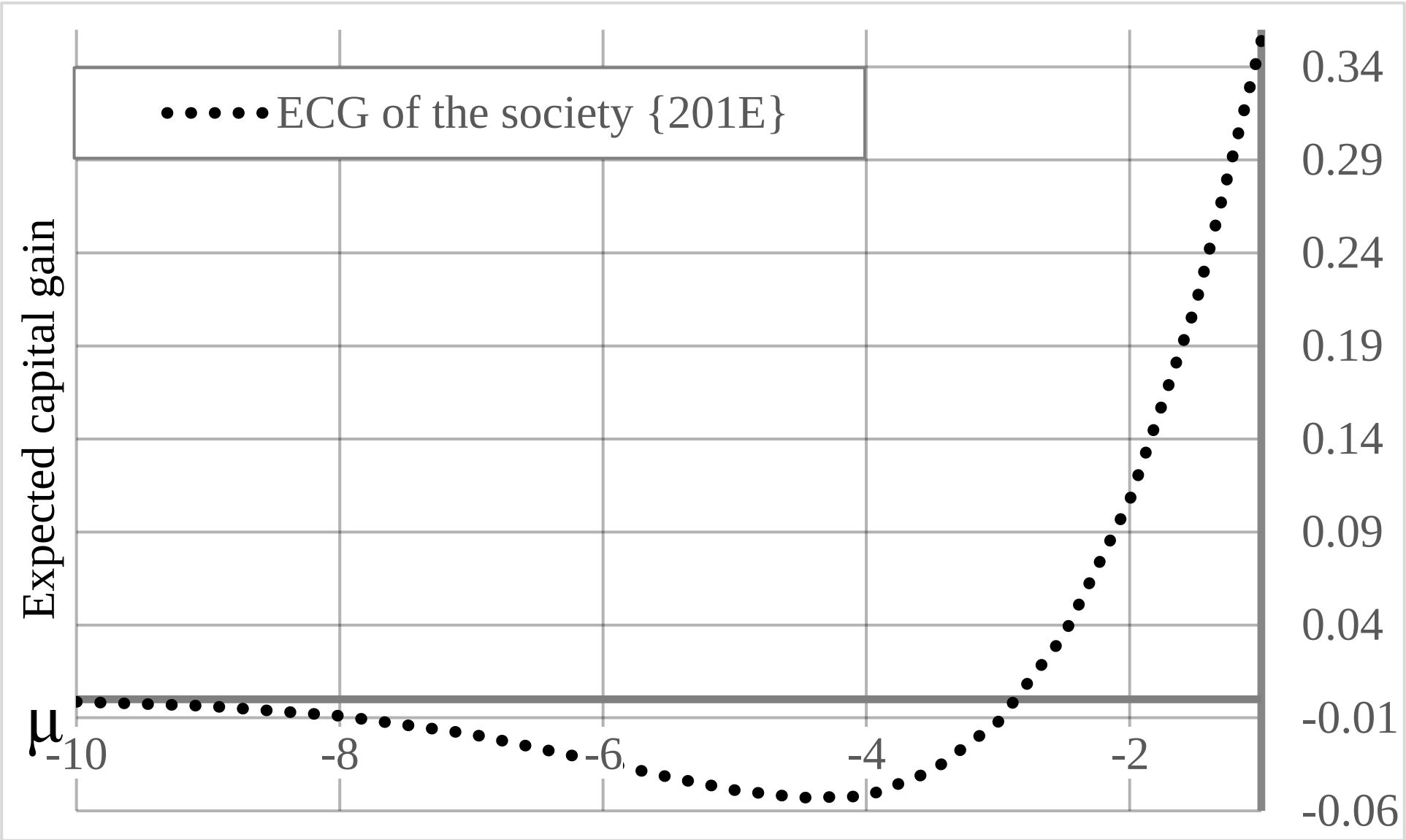

Fig. 1. Expected capital gain (ECG), 201 agents (egoists), $\sigma = 34$.

Fig. 2 shows the ECG curves for the societies containing 5 or 11 altruists among 201 agents, while the rest of the agents are egoists: {5A, 196E} and {11A, 190E}. For comparison, the ECG curve for the society of egoists is shown. Each altruist supports those, and only those, proposals that increase the total capital of the society.

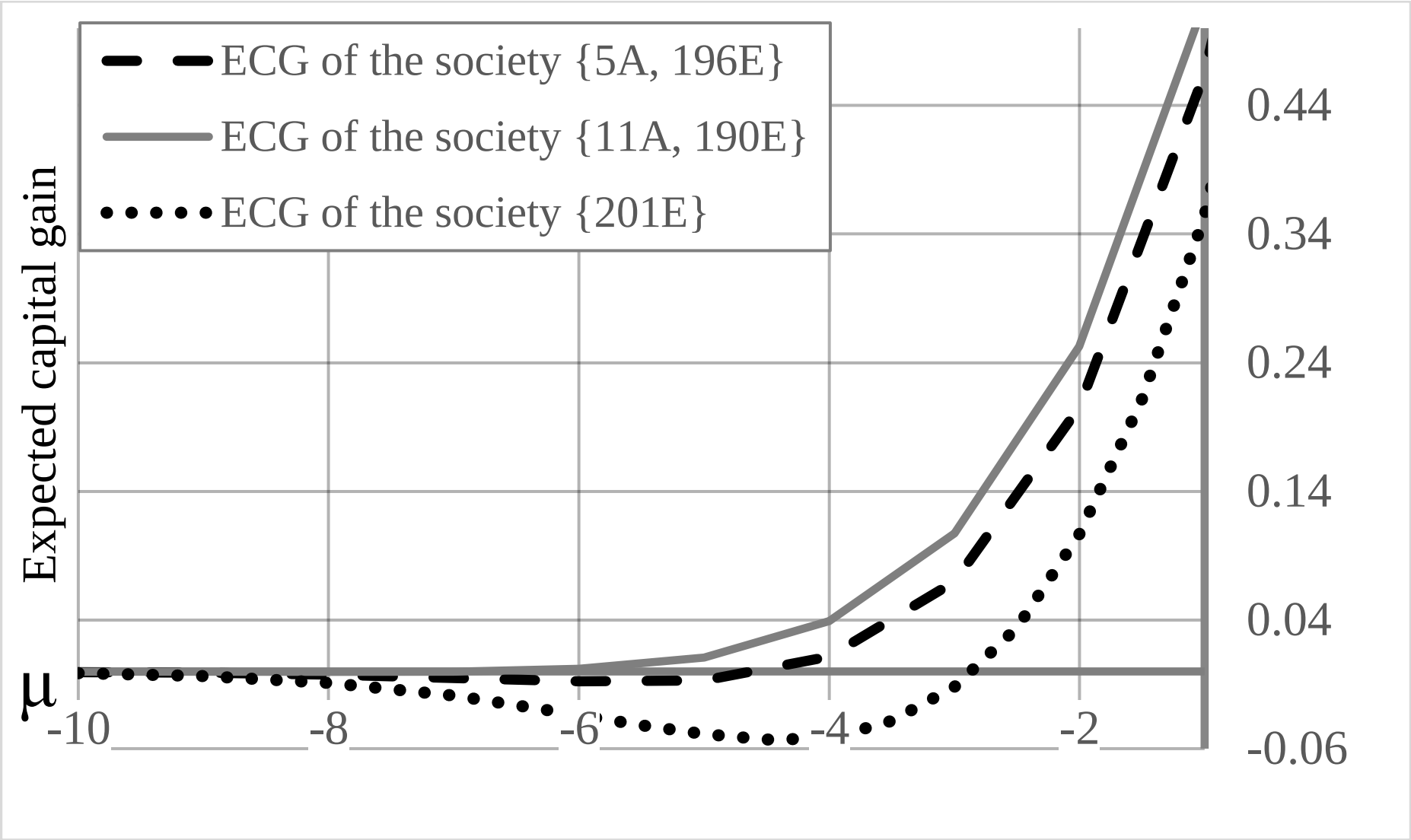

Fig. 2. ECG of the society consisting of egoists (E) and altruists (A).

With five altruists, the 'pit of damage' becomes shallow, and with 11 altruists, it is negligible.

This is achieved at the price of fast impoverishment of altruists. Fig. 3 shows the ECG curves of altruists in the societies under consideration. Five altruists go bust much faster than the agents in the completely selfish society {201E}. However, 11 altruists lose their capital much slower. When the number of altruists is 14, their ECG does not go significantly below zero under any $\mu$.

Thus, if 7% of the agents (14 of 201) agree to vote altruistically, they rescue from bankruptcy not only the whole society (for which 11 members, i.e., 5.5% is enough, see Fig. 2) but also themselves!

On the other hand, the ECG of altruists is significantly lower in these cases than that of egoists in the same society. In Fig. 3, for comparison, the ECG curve of egoists in a society with 11 altruists is shown. This society is denoted by {11A, 190E}. For it, ECG(11A) < ECG(190E). The number of altruists, at which these values are approximately the same, is about 30.

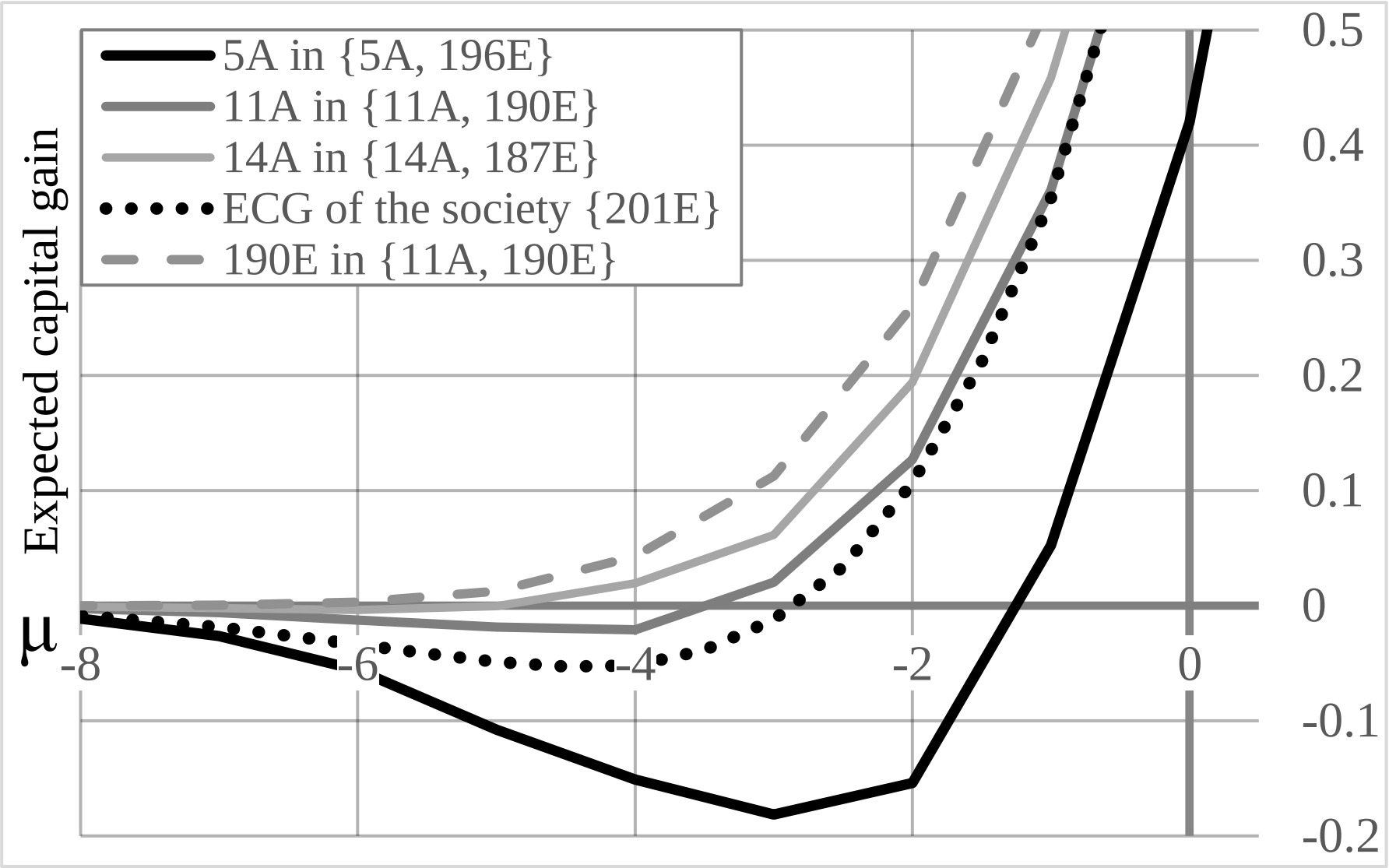

Fig. 3. ECG of the factions in the society consisting of egoists (E) and altruists (A).

In most current societies, the number of altruistic agents who are willing to have low incomes for the sake of others is limited. How is it possible to ensure that these participants receive adequate remuneration, for example, so that their income is not lower than average? However, will this not require investing all the profits that society receives from altruists? The following study answers these questions.

*3.2. Agents with a combined strategy as a responsible elite*

Consider now, instead of altruists, a faction of agents with the following combined strategy. Evaluating a proposal, an agent computes $\alpha D_1 + (1-\alpha)D_2$, where $0 < \alpha < 1$; $D_1$ and $D_2$ being the average capital increments, according to the proposal, of the whole society and of the faction the agent belongs to. The agent supports the proposal whenever $\alpha D_1 + (1-\alpha)D_2 > 0$. Thus, the voting strategy of the agent is based on a convex combination of the altruistic objective function and the *group* one (which is the average capital increment of her group).

If 11 agents of 201 form a faction with such a strategy, and $\alpha = 0.98$, while the rest of the agents are egoists, then the society is denoted by {11(0.98A + 0.02G), 190E}, where A, E, and G designate the altruistic, egoistic, and group strategies, respectively.

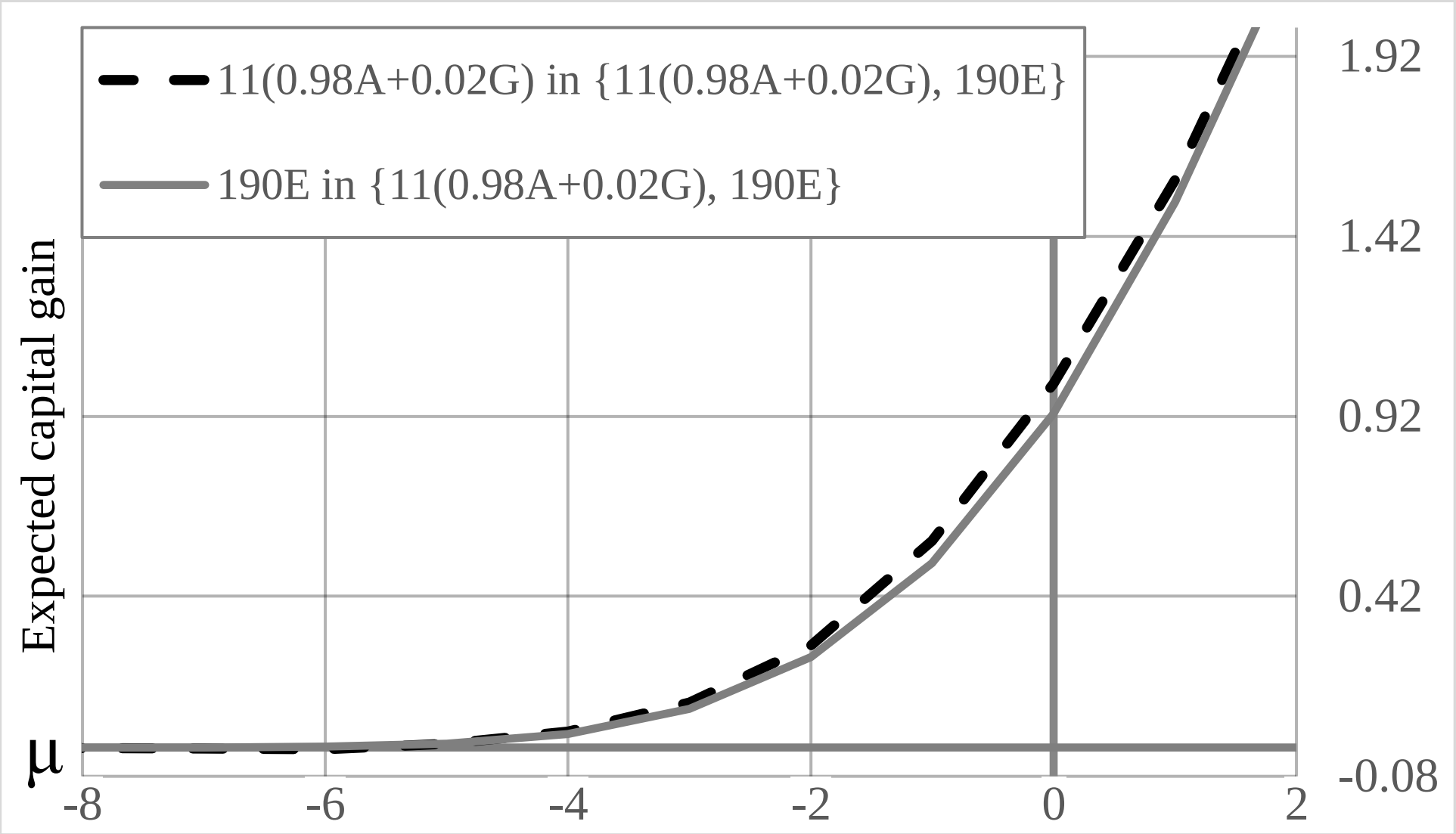

Fig. 4. ECG of the factions in the society, where 11 agents have combined strategy 0.98A+0.02G and 190 agents are egoists.

While the incomes of 11 altruists are significantly lower than those of egoists (Fig. 3) are, the incomes of 11 agents with the above almost-altruistic combined strategy (Fig. 4) are even slightly higher than those of the egoists in the same society are. How do *average* incomes differ in these two societies? As Fig. 5 shows, they are almost equal and significantly higher than those in the society {201E}.

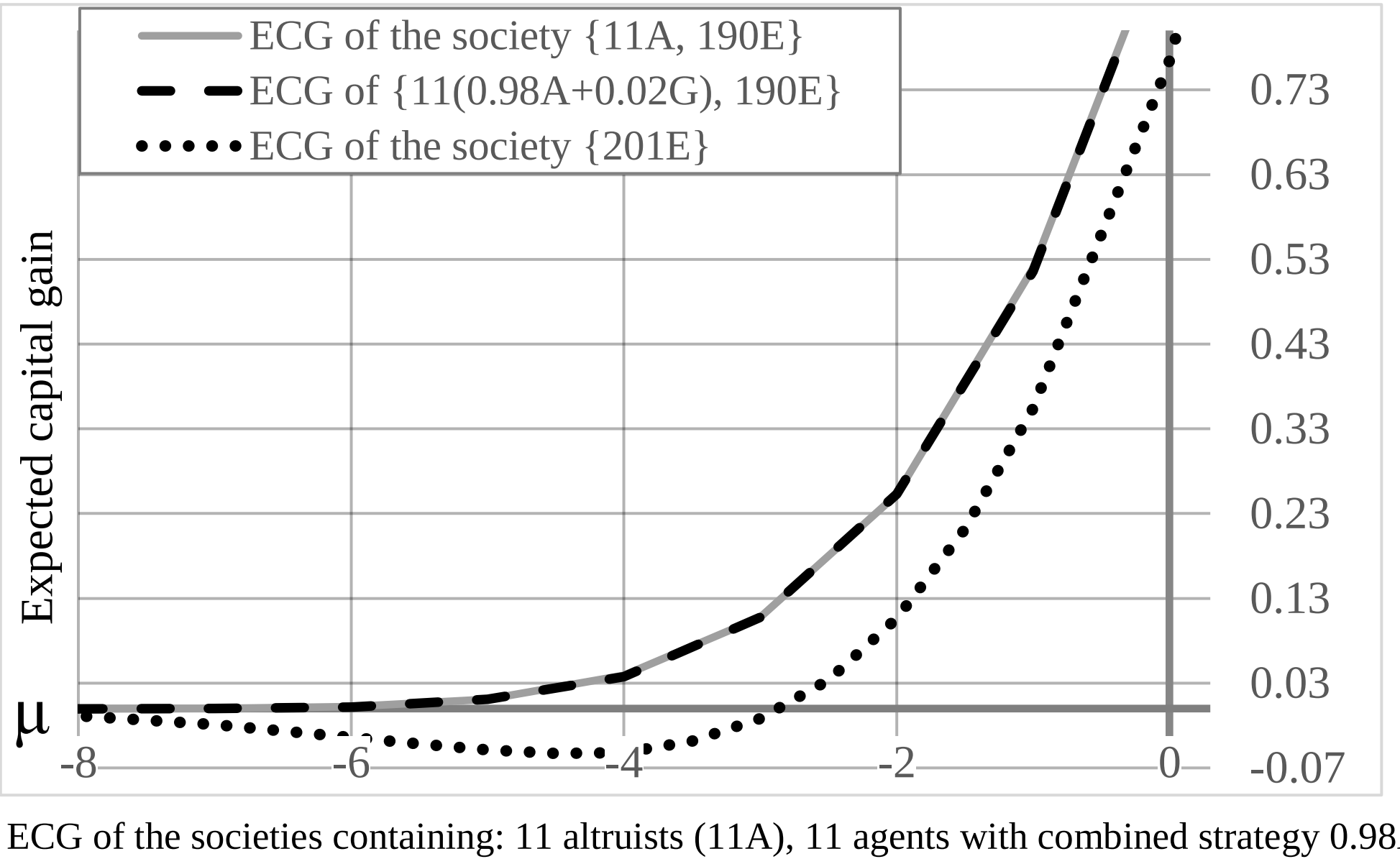

Fig. 5. ECG of the societies containing: 11 altruists (11A), 11 agents with combined strategy 0.98A+0.02G, compared to the society consisting of egoists.

Thus, the combined voting strategies allow the solution of two problems: (1) preventing society from bankruptcy, thanks to a relatively small faction adopting such a strategy; (2) providing members of this faction with an expected income not lower than the average income in the society (Fig. 4). Such a faction can be regarded as a model of *responsible elite.*

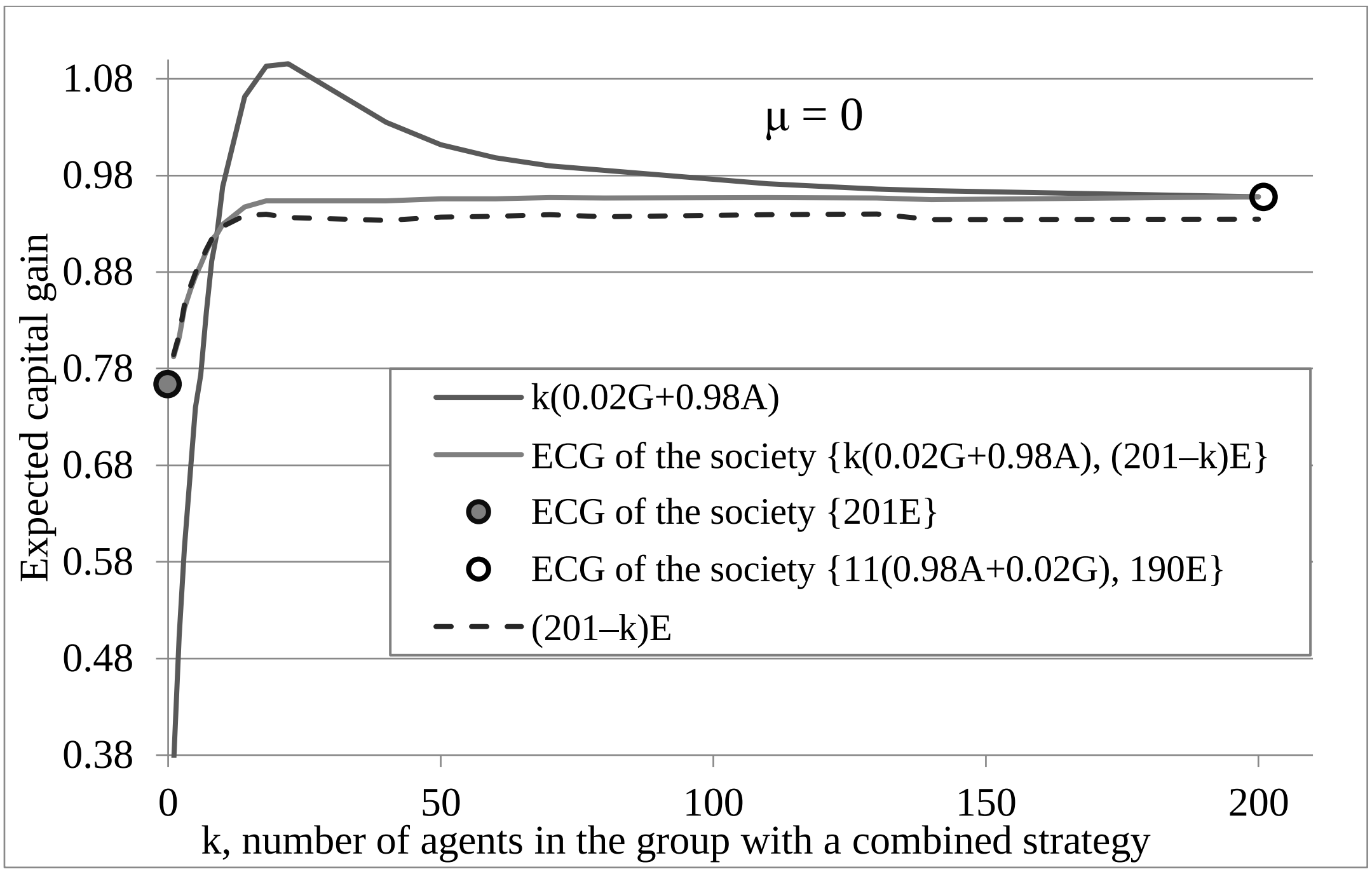

Fig. 6. ECG of the society and its factions when variable number of agents has combined strategy 0.98A+0.02G.

Due to the above properties, joining this faction is beneficial for other agents. What happens in this case? First (the case $\mu = 0$ is shown in Fig. 6), as the size of the faction grows, the incomes of its members grow fast, from very low to high values. With twenty-plus members of the faction, the average income of its members reaches a maximum, and then begins to decrease to the amount of income in the society consisting of altruists; this amount is shown by an open circle. Meanwhile, the incomes of egoists, and of the whole society, grow from the amounts in the society of egoists (shaded circle), and almost stabilize when there are about 15 agents with a combined strategy.

Therefore, for members of the faction with strategy 0.98A+0.02G, a relatively small faction size is most beneficial; here, this size slightly exceeds 10%. The rest of the agents benefit from joining the faction, whereas the size of the faction (exceeding 10%) has little effect on the income of non-affiliated agents.

*3.3. Elite's neglect of social responsibility and reaction to it*

Let us return to the society where 11 agents have strategy 0.98A+0.02G, while the rest of the agents are egoists (Fig. 4 and Fig. 5). The faction of 11 agents protects society from bankruptcy and maintains the incomes of its members at a slightly higher level than the average income in society. The members of this faction can try to enrich themselves by increasing the share of group objective function in their combined strategy. The curve corresponding to this share of 1.0 is shown in Fig. 7. It indicates that this increase really enriches the members of the faction.

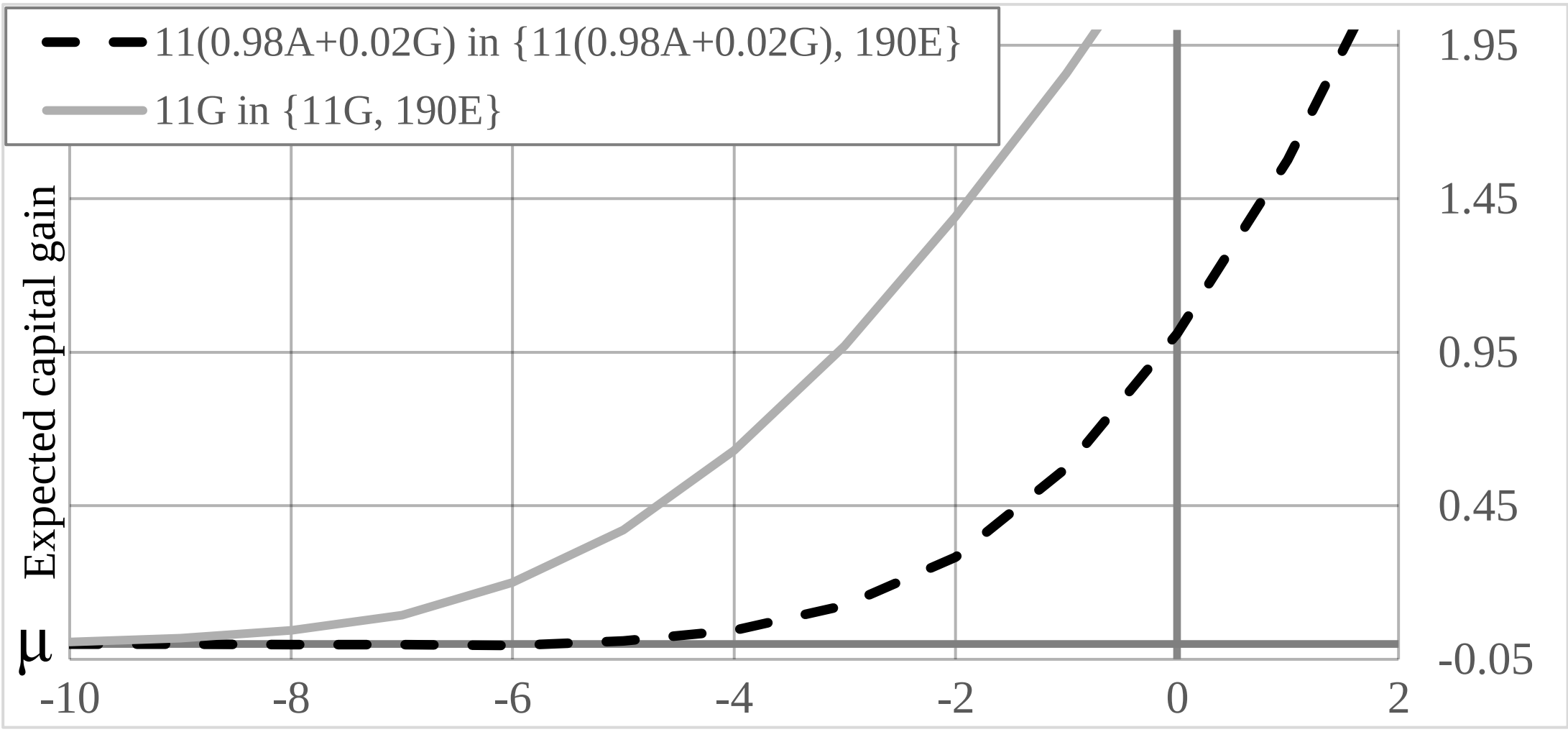

Fig. 7. ECG of 11 agents with the combined strategy 0.98A+0.02G and 11 agents with pure group strategy G.

How does this affect the rest of society? For the rest of society, the elite's neglect of social responsibility and turn to selfishness is extremely disadvantageous. Fig. 8 shows that, in this case, the average agent’s income and the egoist’s income are much lower than the income in a purely egoistic society. Such a faction models the *irresponsible elite*, which may also be called a *clique*.

How can the rest of society respond to this challenge? The remaining agents can, among other reactions:

(1) Join the clique (a ‘snowball of cooperation’ [12]);

(2) Form a new faction with a group strategy [6];

(3) Form a new faction with an altruistic or combined strategy.

Note that, with the successive entry of egoists into the clique, the society reaches the average incomes of a purely egoistic society only when the group consists of 128 agents (64% of the society).

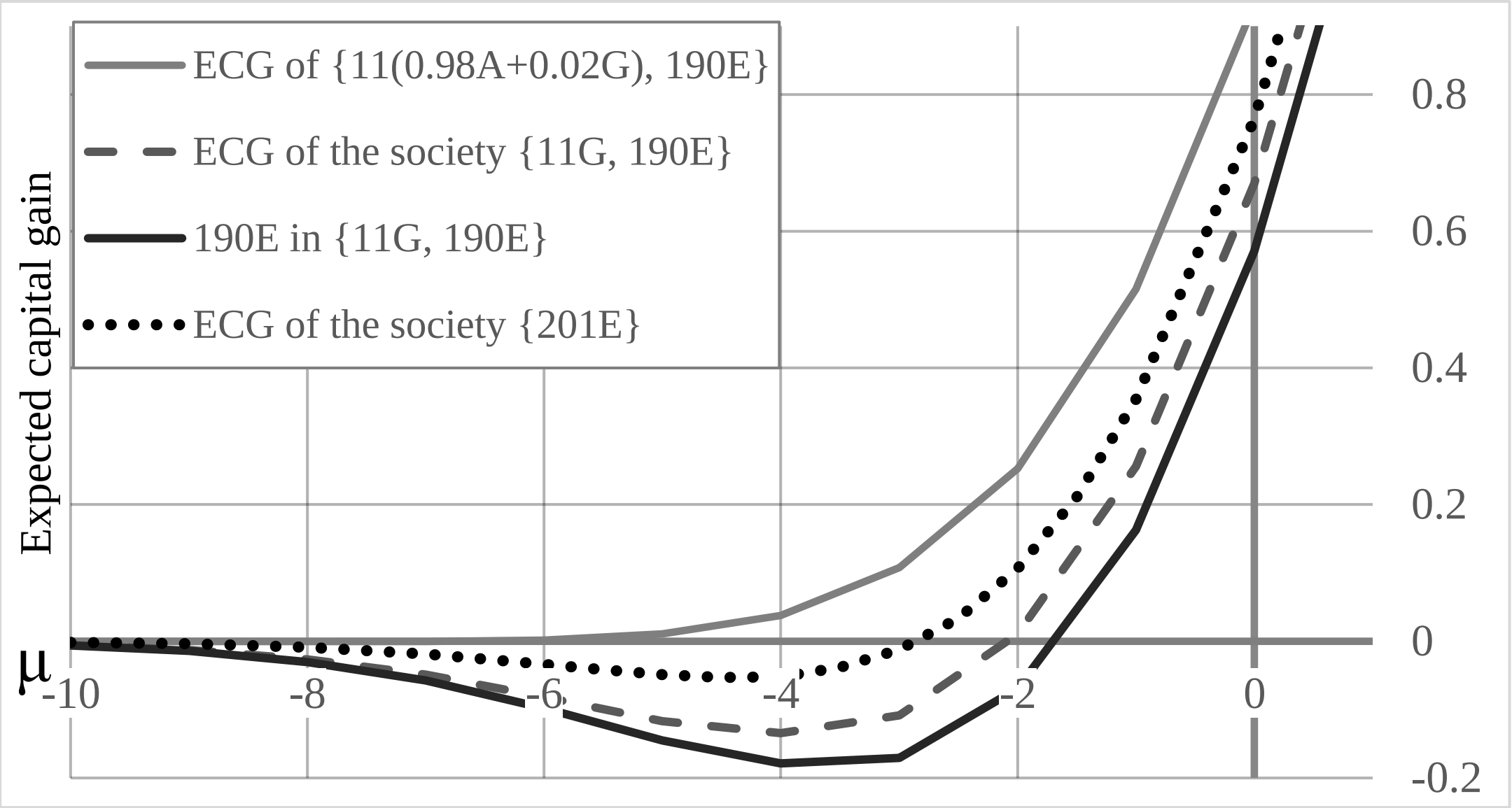

Fig. 8. Impact of the transition of a group of 11 agents from the combined strategy to the group one upon the average income of society members and the egoists.

With further growth of the clique, the income of an egoist and the average income in the society increase, but the income of a clique member decreases, still remaining higher than the income of an egoist (Fig. 9). Finally, the clique that coincides with the whole society is identical to the society of altruists.

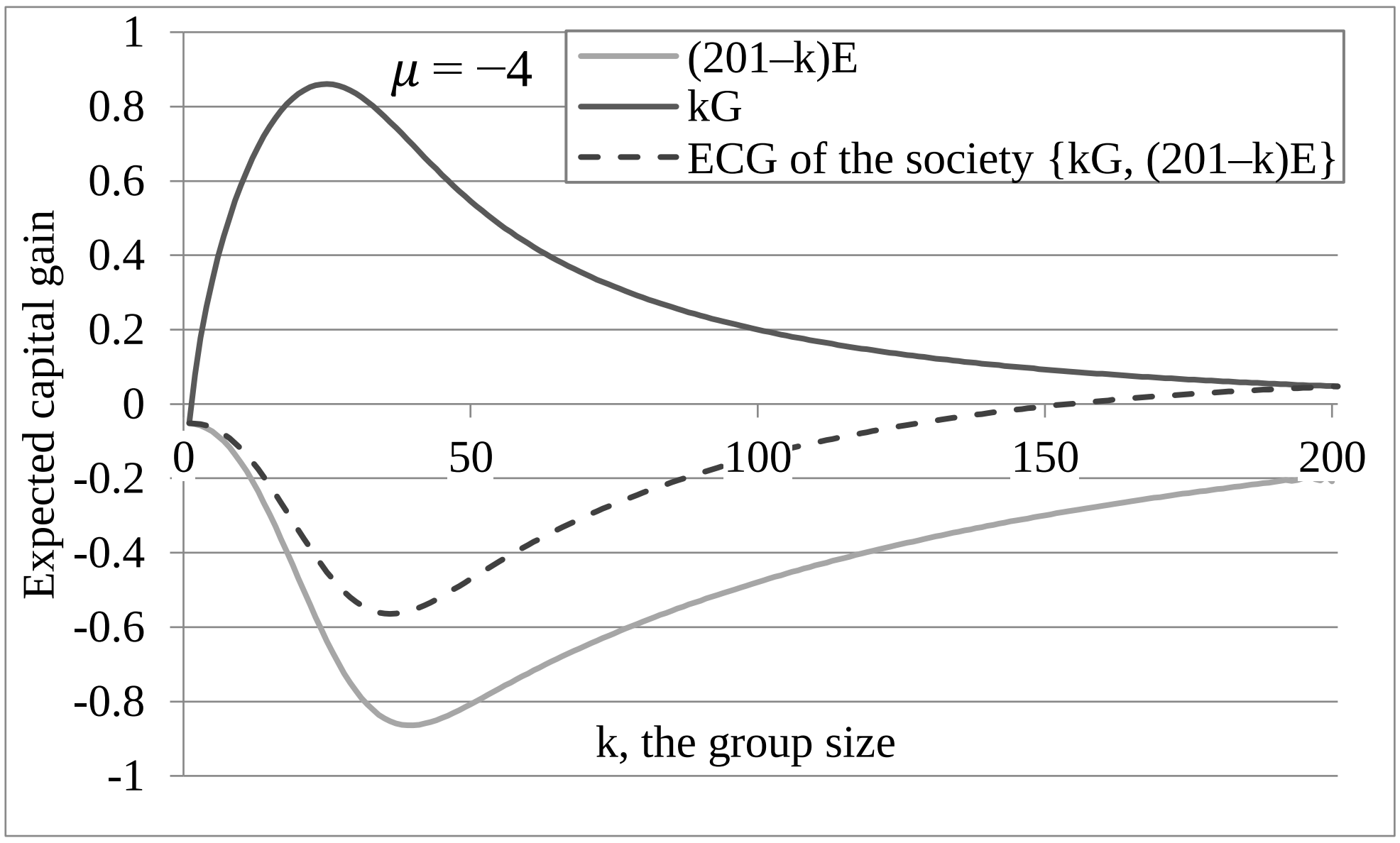

Fig. 9. ECG of the group, the egoists, and the general ECG as functions of the size of the group.

In what follows, we consider the third, asymmetric response to the appearance of a clique, i.e. ‘form a new faction with an altruistic or combined strategy’.

Note (Fig. 10) that the appearance of only five altruists compensates the society for the presence of the selfish group (clique) of 11 agents. The society's ECG reaches the same value as in the egoistic society: ECG({11G, 5A, 185E}) ≈ ECG(201E). It is definitely a significant improvement compared to the society consisting of a clique and egoists.

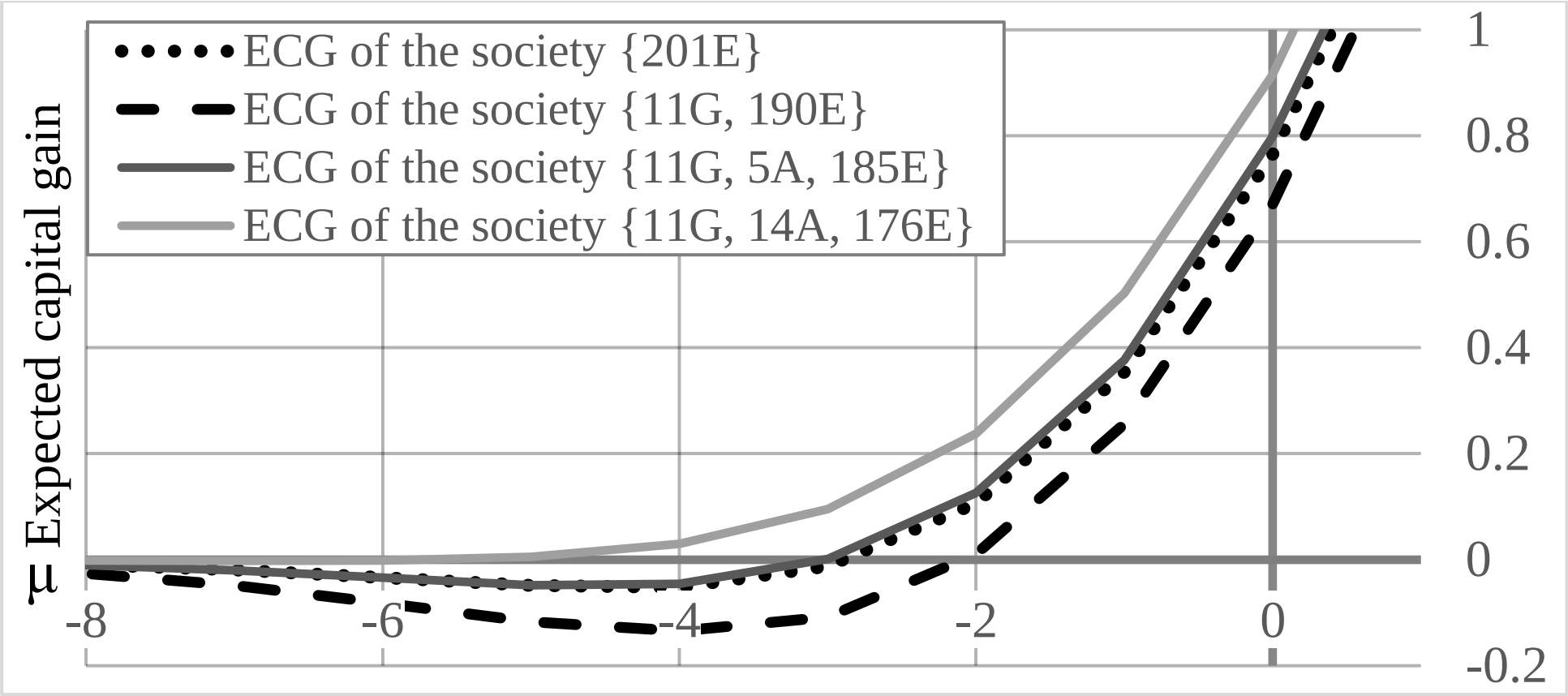

Fig. 10. The impact of the emergence of altruists on the ECG in a society with a clique (group of 11 agents).

When the number of altruists reaches 14, the society stops getting poorer thanks to its collective decisions. However, as usual, the income of altruists is inferior to that of egoists (Fig. 11).

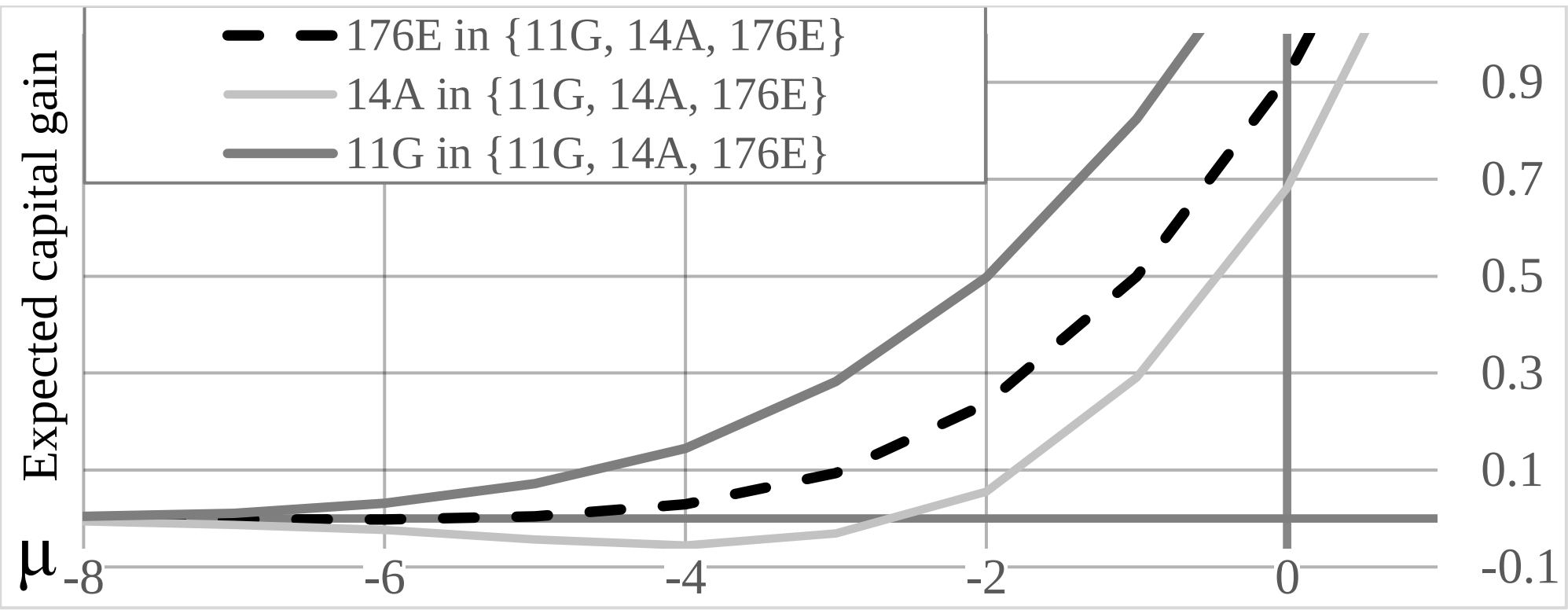

Fig. 11. ECG of the factions in a society consisting of altruists, a clique, and egoists.

With a further increase in the number of altruists, the ECG's of egoists, society, and altruists themselves, grow, whereas the ECG of the clique decreases quite rapidly (Fig. 12).

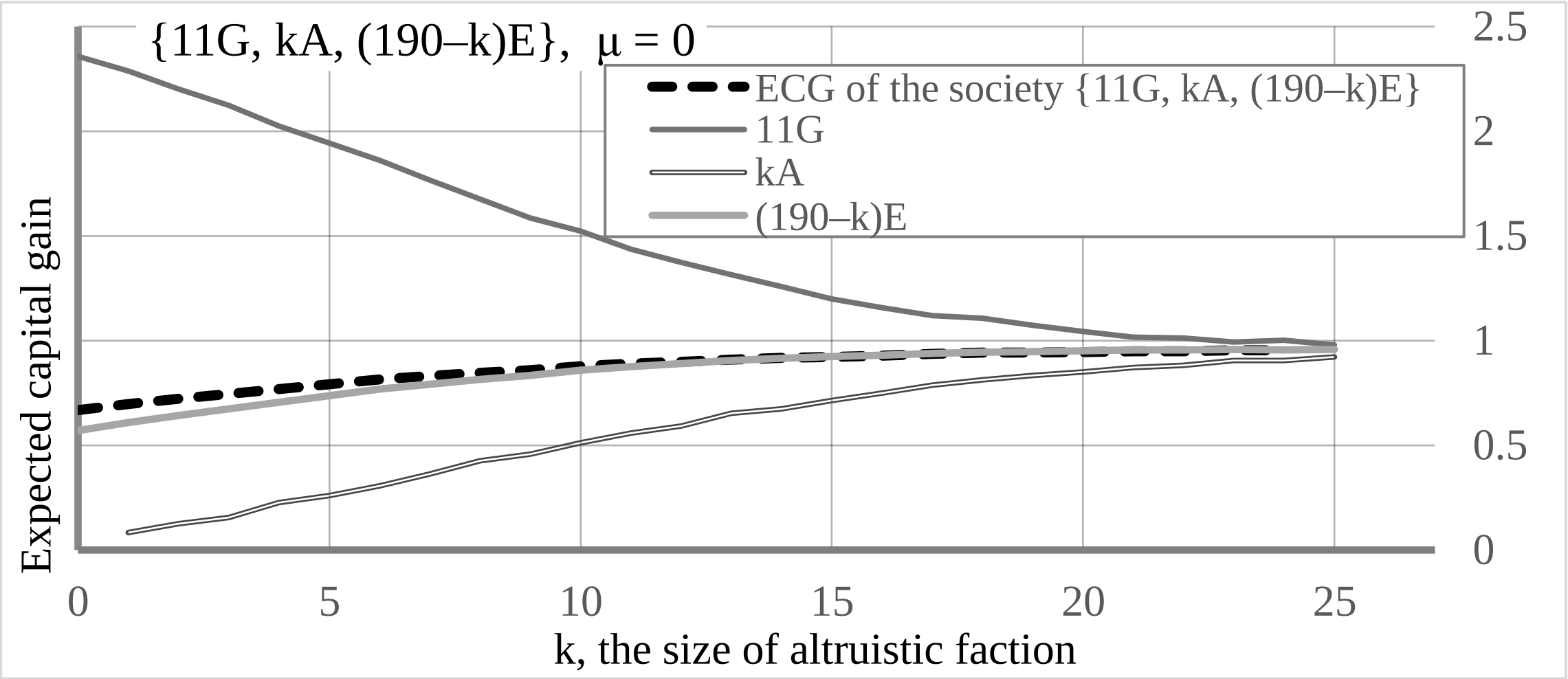

Fig. 12. ECG in the society {11G, kA, (190–k)E} as a function of the number of altruists.

Now, let altruists partially support themselves, by adopting a combined voting strategy. Fig. 13 shows the ECG curves of the whole society, when the percentage of the group's objective function in the combined strategy of former altruists is 4% or 13%, compared to the societies considered earlier. The difference between the curves of 4% and 13% is not very big.

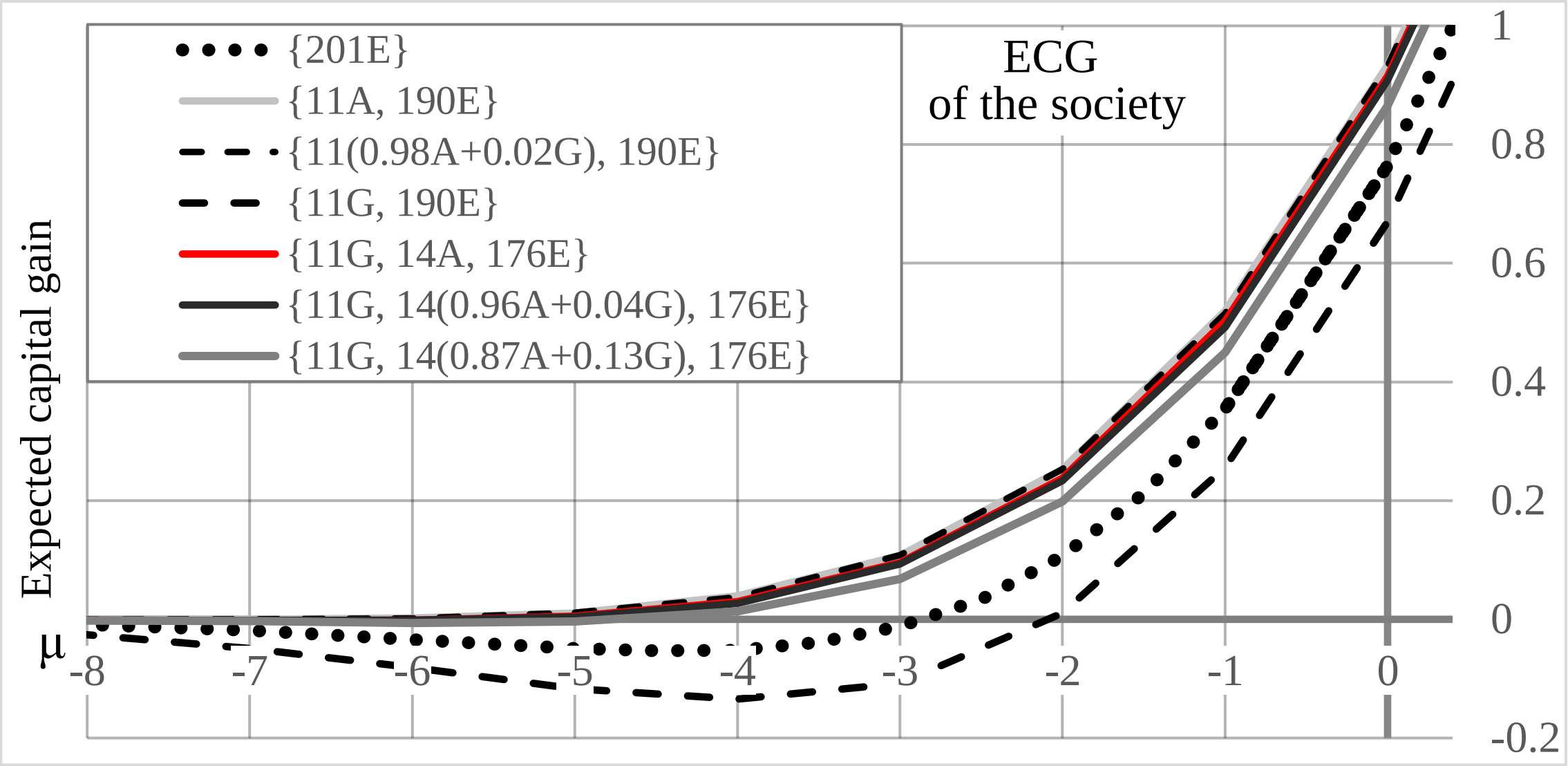

Fig. 13. ECG of societies consisting of a clique, a faction with a combined strategy, and egoists, compared to other societies.

The ECG curves of the society {11G, 14(0.96A+0.04G), 176E} and of its factions are presented in Fig. 14. Here, 14 agents with a combined strategy (a responsible elite) have above-average incomes, which makes their social function not too sacrificial.

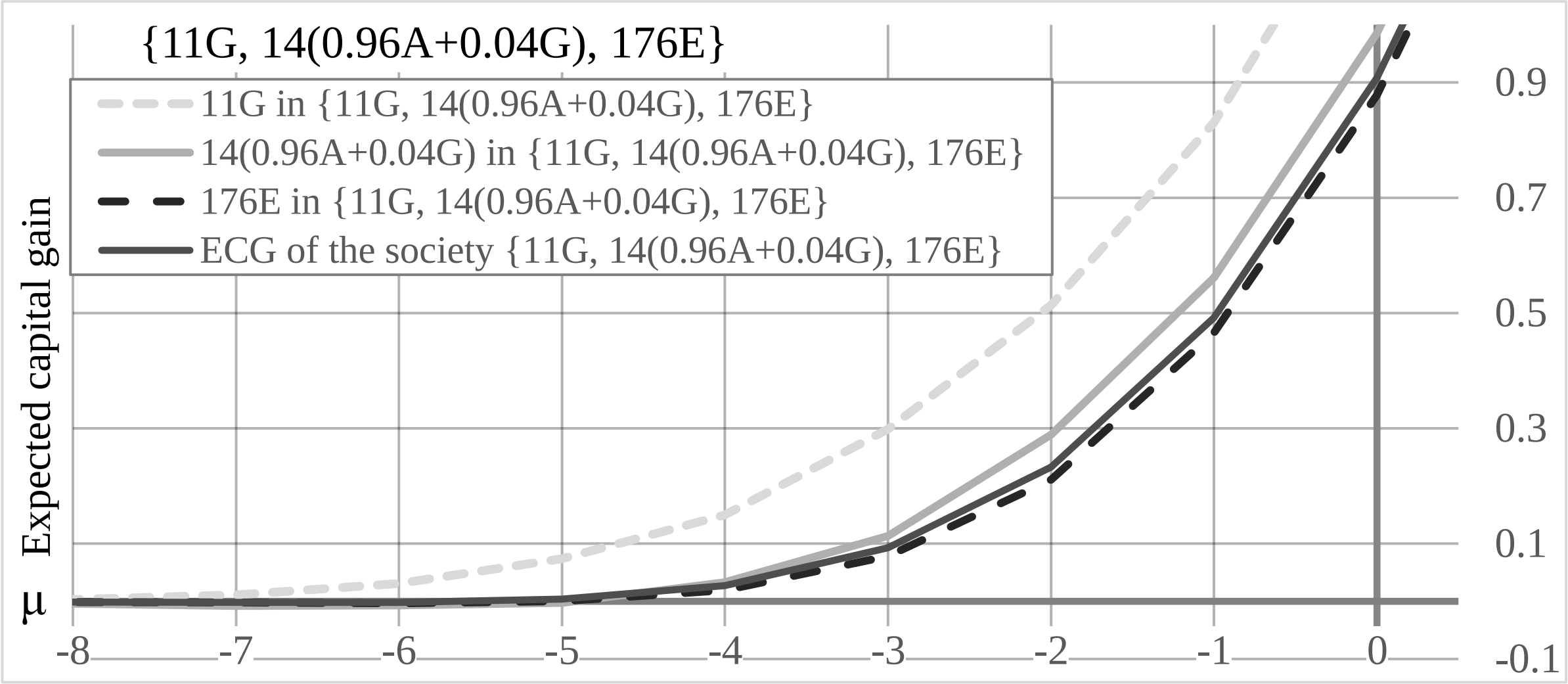

Fig. 14. ECG of the factions in the presence of a clique and 4% of the group objective function in the combined strategy.

It is important to note that, if the faction with a combined strategy increases the proportion of the group objective function to 13% (Fig. 15), then its income overtakes the income of the clique! In this case, the society as a whole does not go bankrupt; its ECG curve has no pit of damage.

Therefore, in the society consisting of a clique (11 agents), a responsible elite (14 agents) and egoists (176 agents), the responsible elite with a fairly low proportion of group strategy (13%) is ahead of the clique and removes society from the zone of imminent bankruptcy, providing it with a non-negative expected capital gain.

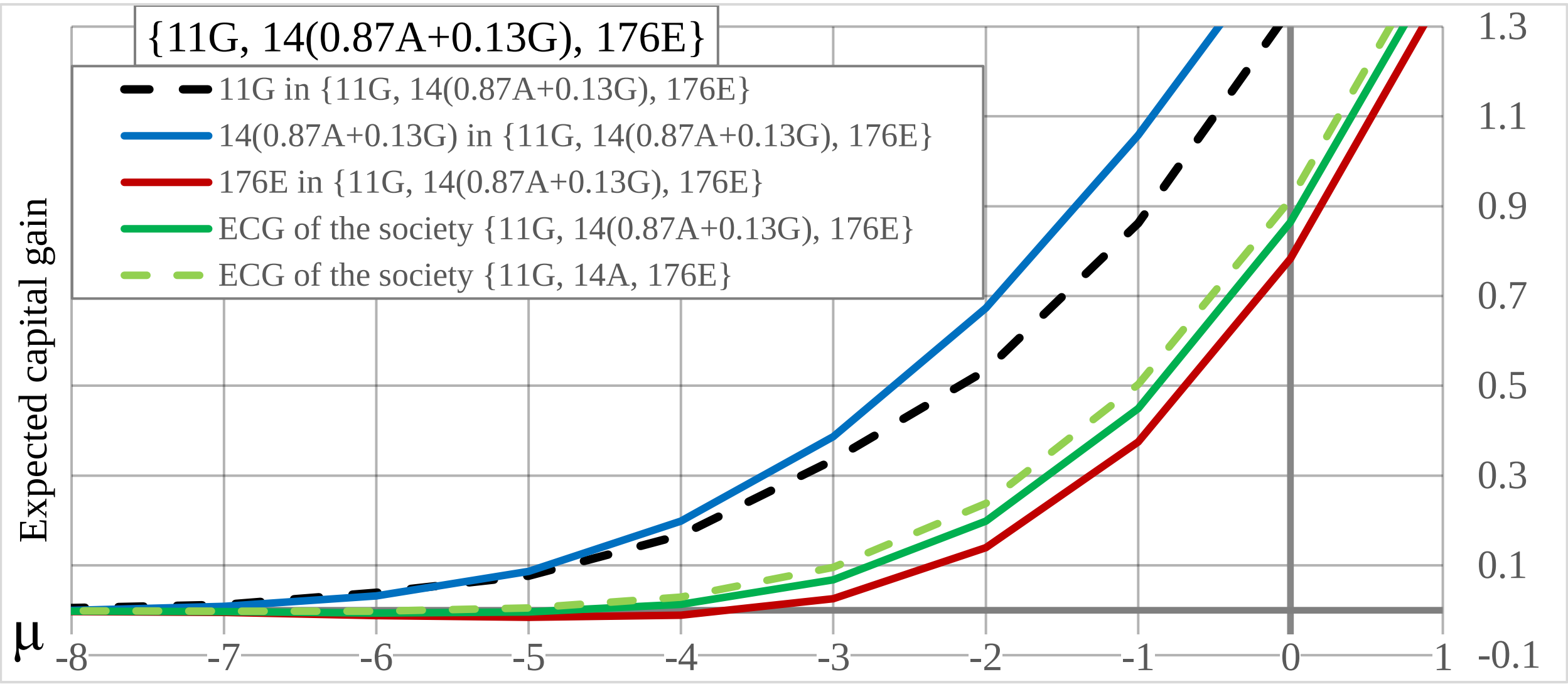

Fig. 15. ECGs in the society with a clique, and a faction whose combined strategy has 13% of the group objective function, compared to a society with an altruistic faction.

Thus, using the ViSE model, we found that a responsible elite, which adheres to a combined voting strategy (allowing it to perform an important social function and take care of itself to the necessary extent), can provide stability for society and help it respond adequately to external challenges of a stochastic nature.

*3.4. Iterative social dynamics*

Finally, let us explore what will happen to society if the second responsible elite becomes a clique like the first one. It has been observed in [3] that tough competition of two cliques is preferable to society over the presence of only one clique. However, if there are two cliques with 11 and 14 agents, the ECG of the egoists and the average ECG in the society are negative (Fig. 16).

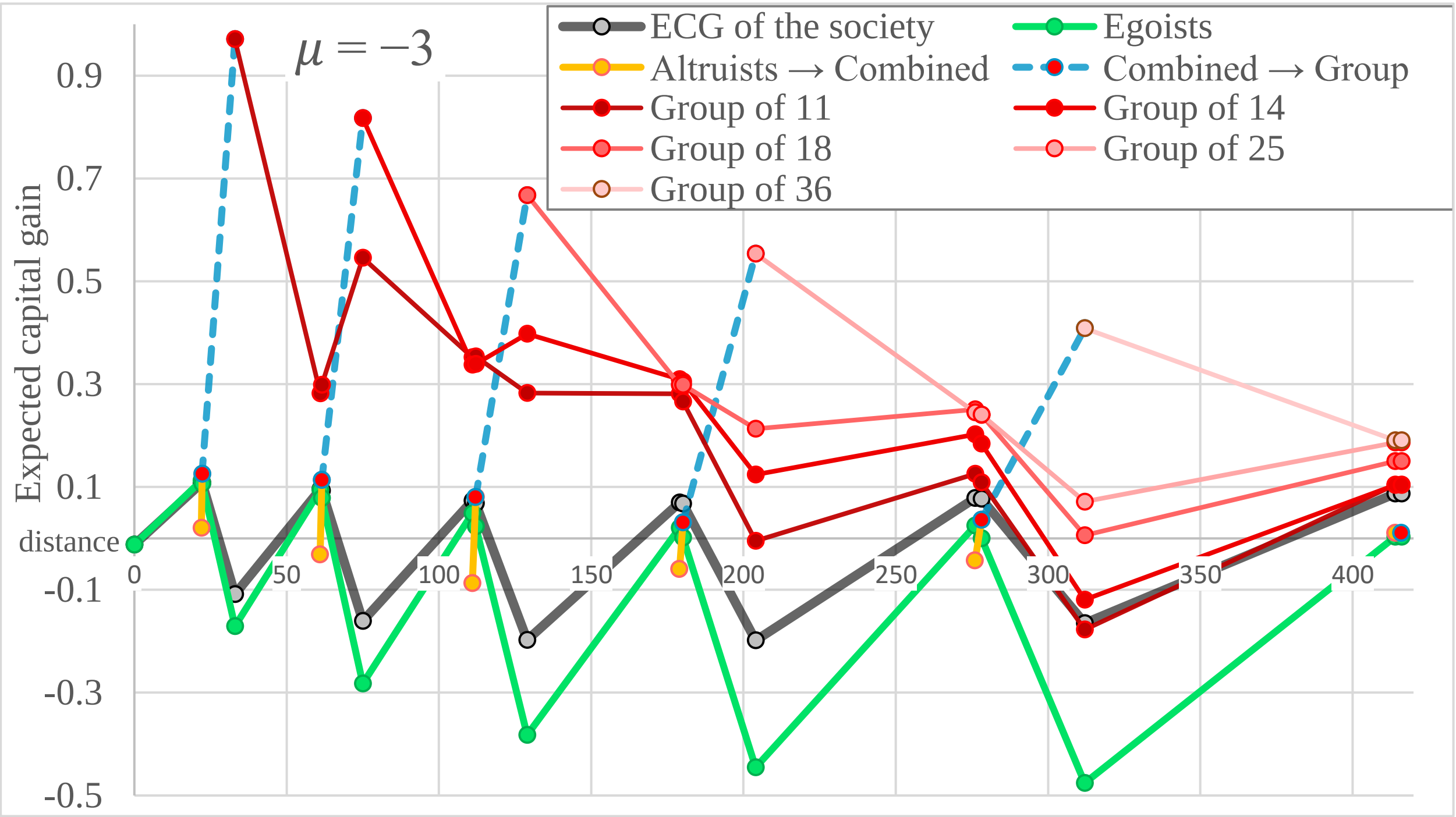

Fig. 16. A sequence of responsible elites that arise in response to the transformation of previous elites into cliques.

As a possible reaction, a new altruistic faction can emerge, which may then turn into a responsible elite. If there are 18 agents in it, then (1) the ECG of the society (and even the ECG of egoists) becomes positive again, and (2) the new responsible elite with an appropriate share of group objective function in its combined strategy can outstrip the ECG of other factions. Fig. 16 presents an *iterative* dynamic of this kind, where the horizontal axis shows a certain distance between societies.

If the third responsible elite also becomes a clique, then another altruistic/responsible faction with the same properties should contain 25 agents. In the next iteration, an altruistic/responsible faction must contain 36 agents, after which, since the society is finite, another responsible elite with the above properties (1) and (2) cannot stand out. It should be noted that in this case, more than half of the society is in different selfish cliques, so that saving all egoists from bankruptcy and exceeding the income levels of all cliques becomes an impossible task.

The list of societies depicted in Fig. 16 is as follows:

{201E};
{11A, 190E};
{11(0.98A+0.02G), 190E};
{11G, 190E};
{11G, 14A, 176E};
{11G, 14(0.96A+0.04G), 176E};
{11G, 14G, 176E};
{11G, 14G, 18A, 158E};
{11G, 14G, 18(0.94A+0.06G), 158E};
{11G, 14G, 18G, 158E};
{11G, 14G, 18G, 25A, 133E};
{11G, 14G, 18G, 25(0.955A+0.045G), 133E};
{11G, 14G, 18G, 25G, 133E};
{11G, 14G, 18G, 25G, 36A, 97E};
{11G, 14G, 18G, 25G, 36(0.94A+0.06G), 97E};
{11G, 14G, 18G, 25G, 36G, 97E}.

Every time a new altruistic faction emerges, the whole society and egoists win; they lose little when this faction becomes a responsible elite with a small share of the group objective function in its combined voting strategy, and they lose a lot when this faction becomes a clique. A clique typically loses from the emergence of a new clique; however, it can gain if these two cliques can enter a new decisive coalition.

## 4. Conclusions

1. The presence of a relatively small responsible elite in society stabilizes it, eliminating the ‘pit of damage’ paradox. The gain of society from the presence of a responsible elite is not very different from the gain, if there is a faction of altruists of the same size.
2. The latter is explained by the fact that, providing the responsible elite with an income slightly higher than the average level of society, requires a fairly small share of the group strategy in the combined voting strategy of this faction.
3. If the responsible elite succumbs to the temptation to radically increase the weight of the group objective function in its combined voting strategy, its income rises sharply, while the income of the entire society decreases. In this case, other agents benefit from joining the elite, but it is beneficial for the elite to maintain its size at a low level, which provides its members with a maximum income.
4. The elite with a high share of group strategy ceases to be ‘responsible’ and protect society, moreover, it impoverishes society. Consequently, there is a growing need for altruistic agents in society. If in counterweight to the formed elite-clique, society forms a responsible elite that exceeds the first one, then this responsible elite manages to stabilize the society, ensure its own income slightly exceeding the average level of society, and significantly reduce the income of the clique. With a sufficient size, the responsible elite is ahead of the income of the clique.
5. If the responsible elite competing with the clique forms a second clique, then a tough competition between two cliques is preferable to the society over having only one clique (see [3]). Such an iterative social dynamic can be continued so that the possibility of the emergence of a responsible elite that ‘saves’ society and is not selfless remains on a series of stages. In light of this, the dynamics under study can be viewed as a struggle between responsible elites and elites-cliques.

The overall conclusion, is that the presence of a responsible elite in society is beneficial to both society and this stratum, however, the responsible elite faces the temptation to reduce its responsibility. The negative effect of this is limited if the elite does not have an exclusive status, that is, if it is possible to form other elites that are willing to abide by a mutually beneficial social contract.

### Acknowledgments

This work was supported by the program of the Presidium of the Russian Academy of Sciences no. 30 “Theory and Technology of Multi-Level Decentralized Group Control in Conditions of Conflict and Cooperation”.

### Bibliography

[1] Borzenko, V.I., Lezina, Z.M., Loginov, A.K., Tsodikova, Ya.Yu., Chebotarev, P.Yu.: Strategies of voting in stochastic environment: Egoism and collectivism, Autom. Remote Control **67(2)**, 311–328 (2006).

[2] Chebotarev, P.Yu.: Analytical expression of the expected values of capital at voting in the stochastic environment, Autom. Remote Control **67(3)**, 480–492 (2006).

[3] Chebotarev, P.Yu., Loginov, A.K., Tsodikova, Ya.Yu., Lezina, Z.M., Borzenko, V.I.: Analysis of collectivism and egoism phenomena within the context of social welfare, Autom. Remote Control **71(6)**, 1196–1207 (2010).

[4] Chebotarev, P.Yu., Malyshev, V.A., Tsodikova, Ya.Yu., Loginov, A.K., Lezina, Z.M., Afonkin, V.A.: The optimal majority threshold as a function of the variation coefficient of the environment, Autom. Remote Control **79(4)**, 725–736 (2018).

[5] Malyshev, V.A.: Optimal majority threshold in a stochastic environment. arXiv preprint arXiv:1901.09233 (2019).

[6] Binmore, K., Eguia, J.X.: Bargaining with outside options, State, Institutions and Democracy. Contributions of Political Economy, Ed. by Schofield, N., Caballero, G., Springer. pp. 3–16 (2017).

[7] Dziuda, W., Loeper, A.: Dynamic collective choice with endogenous status quo, J. Polit. Economy **124(4)**, 1148–1186 (2016).

[8] Penn, E.M.: A model of farsighted voting, Amer. J. Polit. Sci. **53(1)**, 36–54 (2009).

[9] Dziuda, W., Loeper, A.: voting rules in a changing environment, SSRN paper 2500777. DOI: 10.2139/ssrn.2500777 (2015).

[10] Hortala-Vallve, R.: Qualitative voting, J. Theoret. Polit. **24(4)**, 526–554 (2012).

[11] Chebotarev, P.Yu., Loginov, A.K., Tsodikova, Ya.Yu., Lezina, Z.M., Borzenko, V.I.: A snowball of cooperation and snowball communism. Proc. Fourth Int. Conf. on Control Sciences, Moscow: Inst. Probl. Upravlen., pp. 687–699 (2009).